\documentclass[a4paper]{jpconf}
\usepackage{graphicx}

\usepackage{amsmath}
\usepackage{amssymb}
\usepackage{graphicx}
\usepackage{dsfont}
\usepackage{dcolumn}
\usepackage{units}
\usepackage{wasysym}
\usepackage{multirow}
\usepackage{slashed}
\usepackage{sidecap}
\usepackage{cite}
\usepackage[usenames]{color}
\usepackage[
colorlinks=true,
pdfencoding=auto,
psdextra,
linkcolor=blue,
citecolor=blue,
urlcolor=blue
]{hyperref}

\usepackage{slashed}
\usepackage{booktabs}
\usepackage{wasysym}

\begin{document}
\title{Electromagnetic decays of the neutral pion investigated in the Dyson-Schwinger formalism }

\author{Esther~Weil$^{1}$, Gernot~Eichmann$^{2}$, Christian~S.~Fischer$^1$, Richard~Williams$^1$}

\address{$^{1}$ Institut f\"ur Theoretische Physik, Justus-Liebig Universit\"at Gie{\ss}en, 35392 Gie{\ss}en, Germany}
\address{$^{2}$ CFTP, Instituto Superior T\'ecnico, Universidade de Lisboa, 1049-001 Lisboa, Portugal}
\ead{Esther.D.Weil@physik.uni-giessen.de}

\begin{abstract}
We summarize recent work in determining the transition form factor (TFF) of the neutral pion ($\pi^0 \to \gamma^*\gamma^*$), by solving the non-perturbative Dyson-Schwinger and Bethe-Salpeter equations. We first study the transition form factor, followed by the rare decay $\pi^0 \to  e^+ e^- $, which requires the TFF as input. In addition to the aspects of truncation required to compute the solution, we discuss unexpected behavior in the large $Q^2$ regime (with $Q^2$ the photon virtuality), and also touch on a path deformation in the complex plane to access the total decay rate for the rare decay.

\end{abstract}

\section{Introduction}

Meson transition form factors are intermediate states of many decay and collision processes. As experiments continue to reduce statistical uncertainties on such observables, it is important to match the effort on the theoretical side. To calculate these non-perturbative quantities we are using the framework of the Dyson-Schwinger and Bethe-Saltpeter equations. We summarize the calculations described in Refs.~\cite{Eichmann:2017wil, Weil:2017knt}. These two publications are concerned with the calculation of the pion transition form factor $\pi^0 \to \gamma^* \gamma^*$ and further leptonic decays of the pion, such as the rare decay $\pi^0 \to e^+ e^-$, which poses an interesting case due to a discrepancy between theoretical and experimental data.

\section{Transition form factor $\pi^0 \rightarrow \gamma\gamma$}
\begin{figure*}[t]
	\begin{center}
		\includegraphics[width=1 \textwidth]{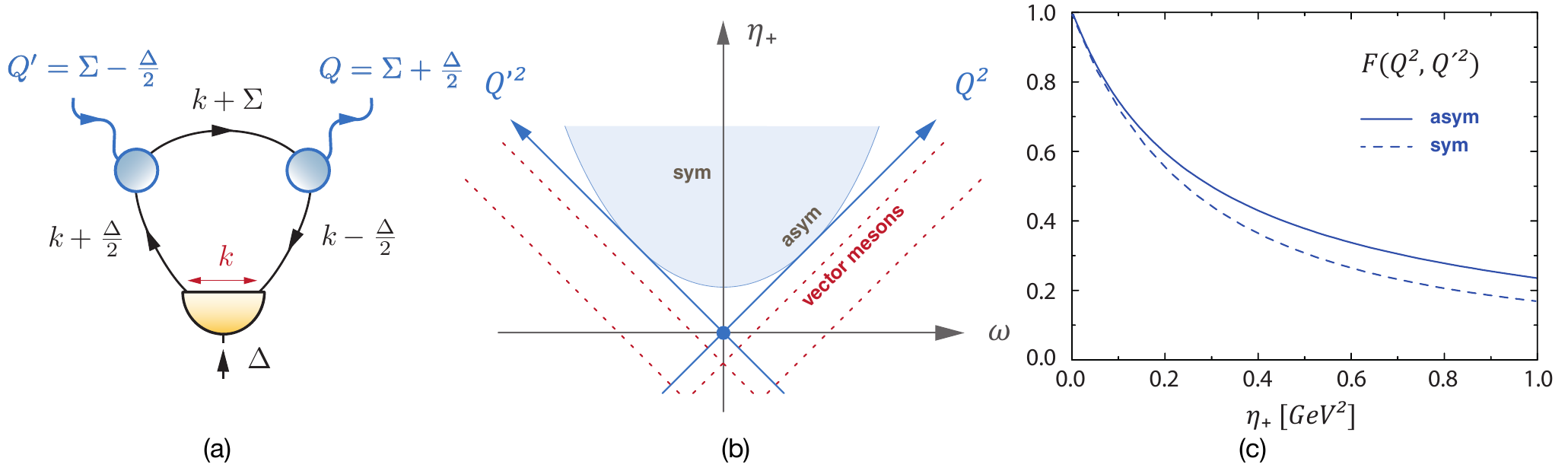}
		\caption{Transition form factor ($\pi^0 \rightarrow \gamma^*\gamma^*$) : (a) transition matrix element, (b) kinematic domains in $Q^2$ and $Q'^2$ including the symmetric and asymmetric limit. The dotted lines indicate the vector pole locations. (c) Result for the on-shell transition form factor in the symmetric and asymmetric limit. Here we use $\omega= (Q^2-Q'^2)/2$  and  $\eta_+ = (Q^2+Q'^2)/2$.
			\label{fig:TFF_all}}
	\end{center}
\end{figure*}

The $\pi\rightarrow \gamma^* \gamma^*$ transition matrix element is given by the tensor
        \begin{equation}\label{pigg-current}
\Lambda^{\mu\nu}(Q,Q') = e^2\,\frac{F(Q^2,{Q'}^2)}{4\pi^2 f_\pi}\,\varepsilon^{\mu\nu\alpha\beta}  {Q'}^\alpha Q^\beta \,,
\end{equation}
 with incoming and outgoing photon momenta $Q'$ and $Q$, the pion's electroweak decay constant $f_\pi\approx 92$~MeV and the
electromagnetic charge $e$. The transition is described by a single scalar function, called a transition form factor (TFF), depending on virtualities of the two photon momenta $F(Q^2,{Q'}^2)$, and the convention of prefactors is chosen such that $F(0,0)=1$  in the chiral limit, constrained by the Abelian anomaly. Besides the on-shell point the form factor is evaluated in different kinematic regions, see Fig~\ref{fig:TFF_all} (b). The so called singly virtual or asymmetric limit (one photon on-shell $Q'^2=0$ and one off-shell  $Q^2\neq 0$) is the quantity measured in most experiments. The symmetric limit ($Q^2=Q'^2$) is the kinematic region needed as an input for the decay into a di-lepton pair $\pi^0 \rightarrow  e^+ e^- $. These are both contained in the space-like domain ($Q^2>0$ ,  $Q'^2>0$). The time-like region ($Q^2<0$ , $Q'^2<0$) is probed e.g when evaluating the pion Dalitz decays $\pi\to \gamma e^+ e^-$,  $\pi^0 \to 2e^+2e^-$. This region contains the physical singularities such as vector meson poles which make the evaluation of the form factor in this region challenging. 

\subsection{The triangle diagram}

In the impulse approximation, the transition form factor $\pi^0 \rightarrow \gamma^{*} \gamma^{*}$ is  given by, Fig.~\ref{fig:TFF_all} (a)
\begin{align} \label{eqn:PseudoScalarFormFactor}
\Lambda^{\mu\nu} = 2e^2\, \text{Tr} \int \!\! \frac{d^4k}{(2\pi)^4} \,  S(k_+)\,\Gamma_\pi(k,\Delta)\,S(k_-)  \Gamma^\mu(r_-,-Q)\,S(k+\Sigma)\,\Gamma^\nu(r_+,Q')\,, 
\end{align}
where $\Sigma= (Q^2+Q'^2 )/2$ is the average photon momentum, $\Delta= Q-Q'$ the pion momentum, $k$ is the loop momentum, $k_\pm = k \pm \Sigma/2$ and $r_\pm = k + \Sigma/2 \pm \Delta/4$. The trace in Eq.~\ref{eqn:PseudoScalarFormFactor} is over Dirac indices only and the factor 2 in front of the diagrams is a symmetry factor. The ingredients of the triangle are obtained by numerically solving Dyson-Schwinger (DSE) and Bethe-Salpeter equations (BSE). 

The internal lines in the triangle diagram Fig.~\ref{fig:TFF_all}(a) are quark propagators $S$. Those include effects of dynamical mass generation due to dynamical breaking of chiral symmetry and we calculate them solving their corresponding DSE. The pion bound-state composed from two quarks is described by the pseudoscalar Bethe-Salpeter amplitude $\Gamma_\pi$ (yellow half circle). Both elements are described by scalar functions times their tensor structure in Dirac, color and flavor space. The quark propagator is described by two tensor structures, while the pion has four. The quark-photon vertex $\Gamma^\mu$ fulfills an inhomogeneous Bethe-Salpeter equation and is decomposed into twelve tensor structures. These can be split into a transverse and non-transverse part, the so-called Ball-Chiu vertex \cite{Eichmann:2016yit}. The transverse part of the vertex contains vector meson states. Thus by properly solving the BSE for the vertex, vector-meson-dominance effects are automatically incorporated into our approach.  

Furthermore it should be mentioned that DSE/BSE's can not be solved without truncating the equations. In context of the form factor calculation we used the so called rainbow-ladder truncation and the Maris-Tandy model \cite{Maris:1999nt} with parameters $\Lambda=0.74$ GeV and $\eta = 1.85 \pm 0.2$. Furthermore the input quark mass is $m_q=3.57$ MeV at a renormalization point of $\mu = 19$ GeV. The pion mass and decay constant we obtain are $m_{\pi^0}=135.0(2)$MeV and $f_{\pi^0} = 92.4(2)$MeV.

\subsection{Results}\label{prev_sec}

With all the building blocks as described in the previous section we were able to determine the form factor in the space-like domain as well as in a small region into the time-like area ($Q^2>-m_\pi^2$). The result for the form factor in the asymmetric and symmetric limit are shown in Fig.~\ref{fig:TFF_all} (c). At the on-shell point we obtain a value of $F(0,0) = 0.996$, which is close to the anomaly constraint of $F(0,0)=1$ in the chiral limit. A direct calculation is however only possible in a restricted kinematic region, because of singularities in the quark propagator. For on-shell pion momentum we are limited to values $Q^2 \lesssim 4$ GeV$^2$ in the asymmetric limit. To calculate the form factor in the whole space-like plane we applied a technique briefly touched up on in section~\ref{section4}. With this we were able to provide a fit for the form factor in the whole space-like domain, which is given in Ref.~\cite{Weil:2017knt}, Eq.~(12). 
 
\section{Rare decay}

\begin{figure}[b]
\includegraphics[width=0.6 \textwidth]{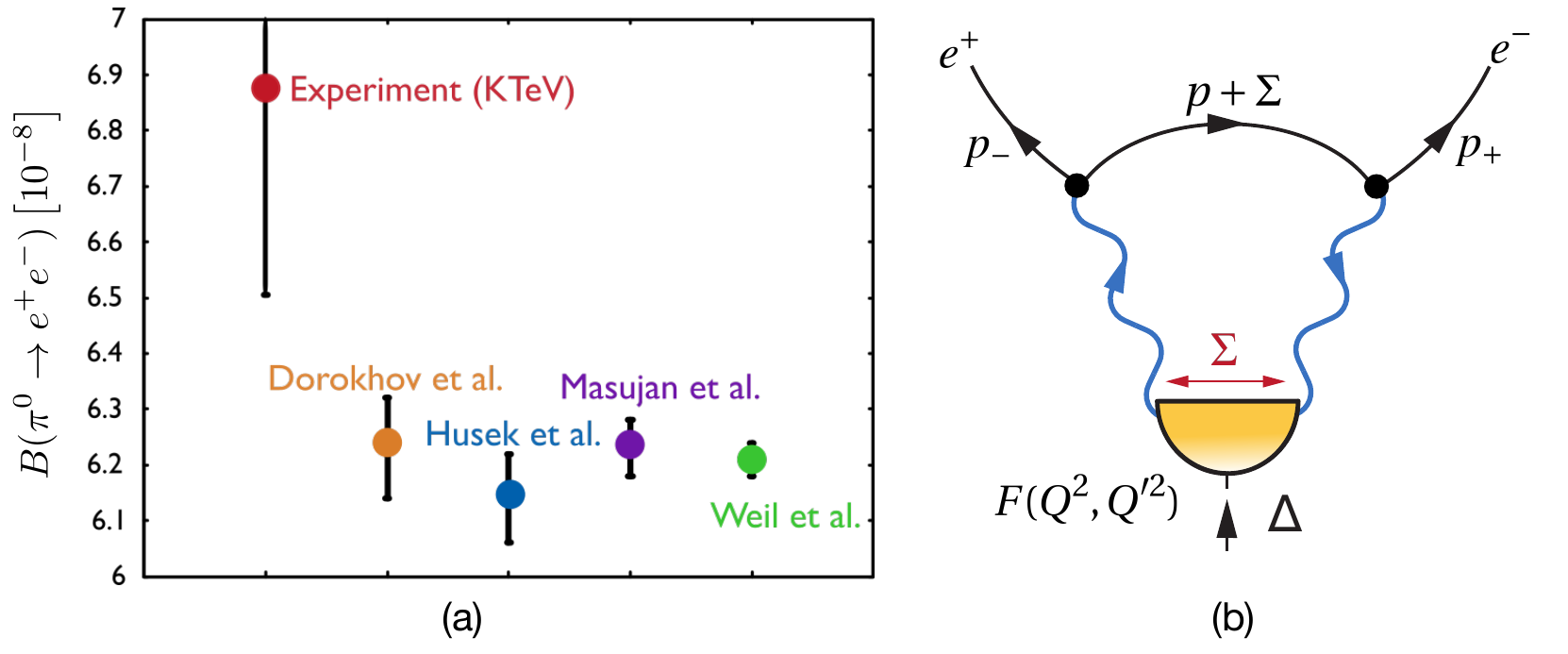}\hspace{0.0pc}%
	\begin{minipage}[b]{14pc}\caption{ Rare decay of the neutral pion $\pi^0\rightarrow e^+ e^- $: 
			(a) Comparison of the latest experimental results \cite{Abouzaid:2006kk,Husek:2014tna} and theoretical calculations \cite{Abouzaid:2006kk,Vasko:2011pi,Husek:2014tna, Dorokhov:2007bd,Dorokhov:2009jd,Husek:2015wta, Masjuan:2015lca}.
			(b) Feynman diagram to Eq.~\ref{A(t)-1}, the yellow half-circle denotes the pion transition form factor $F(Q^2,Q'^2)$, as only non-perturbative input.
			\label{fig:raredecay}}
	
	\end{minipage}
\end{figure}

The two-body decay of the neutral pion into a di-lepton pair $(\pi\rightarrow e^+ e^-  )$ poses an interesting puzzle, as the current theoretical estimates show a discrepancy
with the experimental result from the KTeV E799-II experiment at Fermilab~\cite{Abouzaid:2006kk,Dorokhov:2007bd} at the order of $2 \sigma$. The experimental value for the branching ratio is $B(\pi^0 \rightarrow e^+ e^-) = (6.87 \pm 0.36) \times 10^{-8}$, after reanalysis subtracting radiative corrections \cite{Vasko:2011pi,Husek:2014tna}.

To lowest order in QED the process is described by the one-loop Feynman diagram, Fig.~\ref{fig:raredecay}(b), which includes the transition form factor $F(Q^2,{Q'}^2)$ as the only non-perturbative input (indicated in yellow). The corresponding normalized branching ratio is given by
\begin{align}
R = \frac{B(\pi^0\rightarrow e^+ e^- )}{B(\pi^0\rightarrow \gamma\gamma )}  = 2 \left(\frac{m\,\alpha_\text{em}}{\pi m_\pi}\right)^2 \beta(t_0) \  \vert \mathcal{A}(t_0)\vert^2\,,
\end{align}
with $t= \Delta^2/4 $ ,   $\beta(t)= \sqrt{1+ m^2 /t}$ stems from the two-body phase-space integration and $B(\pi^0\to \gamma\gamma )=0.988$. The scalar amplitude $\mathcal{A}(t)$ is the averaged and spin-summed matrix element, which must be evaluated at the on-shell pion point $t_0 = -m_\pi^2/4$.

After taking traces the result reduces to 
\begin{align}\label{A(t)-1}
\mathcal{A}(t) &= \frac{1}{2\pi^2 t}\int \! d^4\Sigma\,\frac{(\Sigma\cdot\Delta)^2-\Sigma^2 \Delta^2}{(p+\Sigma)^2+m^2}\,\frac{F(Q^2,{Q'}^2)}{Q^2 \,{Q'}^2}\,.
\end{align}
This integral has poles in the integration domain and thus cannot be naively integrated. A common technique to work around problems of such kinds is the use of dispersive methods (see e.g. ~\cite{Bergstrom:1983ay, Dorokhov:2007bd,Dorokhov:2009jd} ). Another way to circumvent the problem is path deformation in the complex plane. For a chosen value of $t$ one knows the locations of poles and branch cuts in the complex plane and can thus find a path to navigate around these infinities, such that the Euclidean integration is possible and not ill-defined. The detailed pole analysis for Eq.~\ref{A(t)-1} and the intricate integration path can be found in \cite{Eichmann:2017wil}, Section IV B. 

We calculated the branching ratio using both methods. For the direct calculation we obtain a branching fraction of $B(\pi^0\rightarrow e^+ e^-) = 6.22(3) \times 10^{-8}$ which compares very nicely to our result reached when using dispersion relation $B(\pi^0\rightarrow e^+ e^-) = 6.21(3) \times 10^{-8}$. We are furthermore in agreement with previous theoretical predictions, see overview Fig.~\ref{fig:raredecay}(a) and thus a $2  \sigma$ discrepancy to the experimental value remains. With this we are able to provide a valuable cross check for previous calculations using dispersive tools \cite{Dorokhov:2007bd,Dorokhov:2009jd,Husek:2015wta, Masjuan:2015lca}.

\section{Large $Q^2$-behavior}\label{section4}

\begin{figure*}[b]
	\begin{center}
		\includegraphics[width=1.0 \textwidth]{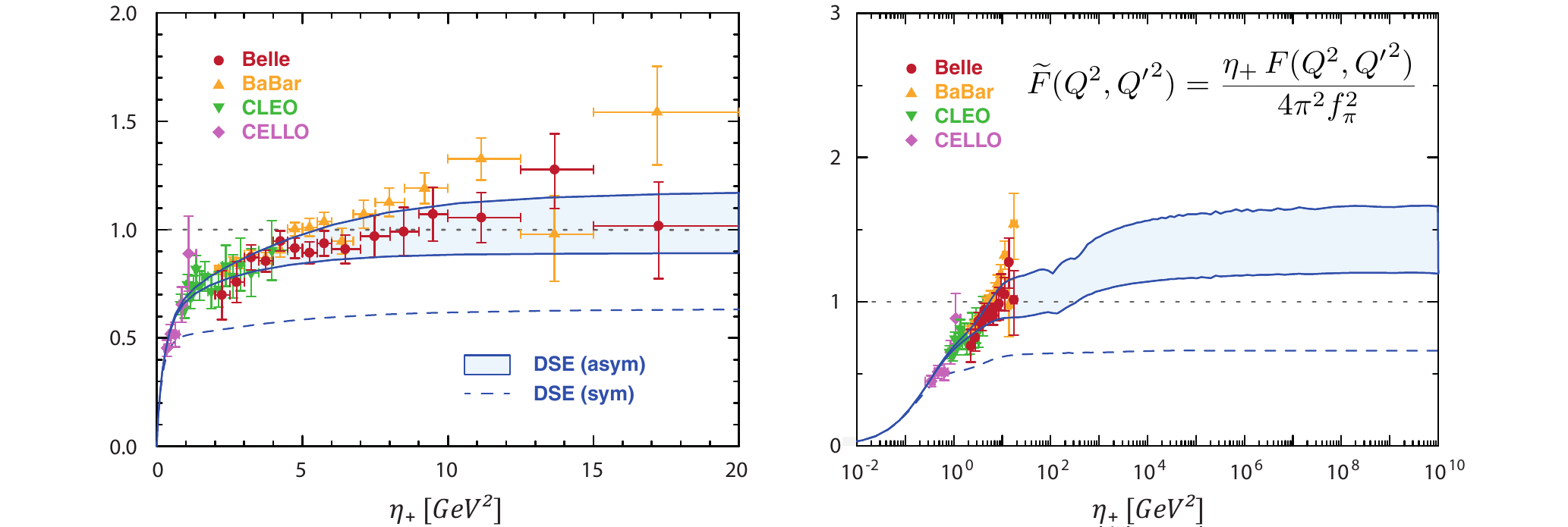}
		\caption{Weighted on-shell transition form factor $\tilde{F}(Q^2, Q'^2)$, as defined in the upper right corner of figure. 
			The blue band shows the result in the asymmetric limit compared to experimental data~\cite{Behrend:1990sr,Gronberg:1997fj,Aubert:2009mc,Uehara:2012ag}, and
			the dashed line is the form factor in the symmetric limit. The error band comes from a variation of fit parameters.
		}\label{fig:ff-largeQ2}
	\end{center}
\end{figure*}

The latest two experimental measurements for the singly-virtual form factor \cite{Aubert:2009mc,Uehara:2012ag} have stirred considerable interest in the large-momentum behavior \cite{Khodjamirian:1997tk,Anikin:1999cx,Melic:2002ij,Radyushkin:2009zg,Polyakov:2009je,Agaev:2010aq,Agaev:2012tm,Arriola:2010aq,Kroll:2010bf,Gorchtein:2011vf,Brodsky:2011yv,Brodsky:2011xx,Noguera:2012aw,ElBennich:2012ij,Dorokhov:2013xpa,Dorokhov:2010zzb,Maris:2002mz,Holl:2005vu,Raya:2015gva,Raya:2016yuj,Mikhailov:2016klg}, with the result from the BaBar collaboration suggesting a deviating from scaling. Within the DSE/BSE framework, the large-momentum behavior of the TFF has been studied previously in Refs.~\cite{Anikin:1999cx,Raya:2015gva,Raya:2016yuj}. We employed a different technique to calculate the form factor in the whole space-like domain and also take into account physical singularities stemming from vector meson dominance, a genuinely non-perturbative effect \cite{Eichmann:2017wil}. 

For off-shell kinematics the form factor is a function of three Lorentz invariants, such as $\{Q^2, Q'^2, Q\cdot Q'\}$, and can be calculated up to arbitrary high space-like values in $Q^2,Q'^2$ without crossing any singularity in the integrand. We thus calculate the off-shell form factor and exploit a fit to the physical point constrained by the vector-meson-pole, see \cite{Eichmann:2017wil} for details. The result for the form factor up the experimentally measured momentum range is displayed in the left panel of Fig.~\ref{fig:ff-largeQ2} and describes the experimental data well. The right panel shows the singly-virtual form factor at asymptotically large momenta beyond the region where experimental data exists. Here we observe a continued rise until the curve plateaus to its asymptotic behavior at around $10^2 - 10^3$ GeV$^2$. The error-band comes from a conservative variation of the fit parameters.  

As compared with the well-know Efremov-Radyushkin-Brodsky-Lepage (ERBL) scaling limit~\cite{Lepage:1980fj,Efremov:1979qk}, which is an asymptotic-QCD prediction, our result agrees in case of the symmetric limit ($Q^2=Q'^2$), but for the asymmetric case exceeds their prediction. We have checked explicitly that discarding the vector-meson pole constraints leads to a recovery of the ERLB limit. The discrepancy still has to be understood. 


\section{Conclusion}

We have summarized recent calculation of the neutral pion TFF using the Dyson-Schwinger and Bethe-Salpeter equations. Generally the results show good consistency with existing theoretical and experimental determinations. 


For the decay into a di-lepton pair ($\pi\rightarrow e^+ e^- $) we confirm the previous theoretical calculation with a branching ratio of $B(\pi^0\rightarrow e^+ e^-) = 6.22(3) \times 10^{-8}$, leaving a $2\sigma$ discrepancy to the experimental value. We were able to provide a cross-check for previous theoretical calculation through using a different method to solve the integral.  

\paragraph{Acknowledgements}
We are grateful to R.~Alkofer, V.~M.~Braun, S.~Brodsky, P.~Kroll, S.~Leu\-pold, M.~T.~Hansen and A.~Szczepaniak for enlightening discussions.
This work was supported by the DFG collaborative research centre TR 16, the BMBF grant 05H15RGKBA,
the DFG Project No. FI 970/11-1, the FCT Investigator Grant IF/00898/2015,
the GSI Helmholtzzentrum fuer Schwerionenforschung, and by the Helmholtz International Center for FAIR.

\section*{References}
\bibliographystyle{iopart-num}
\bibliography{baryonspionff}
\end{document}